\documentclass[journal]{IEEEtran}
\usepackage[pdftex]{graphicx}
\usepackage{subfigure}
\usepackage{balance}
\usepackage{array}
\usepackage{times}
\usepackage{rotating}
\usepackage{multirow}
\usepackage[cmex10]{amsmath}
\usepackage{subfig}
\usepackage{tabularx}
\usepackage{float}
\restylefloat{figure or table}

\begin{document}

\title{Performance Evaluation of Spatial Complementary Code Keying Modulation in MIMO Systems}
\author{Amir H. Jafari, Timothy O'Farrell\\
  Dept. of Electronic \& Electrical Engineering,\\
  University of Sheffield,\\
  Sheffield, United Kingdom, S1 3JD\\
  \texttt{{a.jafari}@sheffield.ac.uk}}
\maketitle

\begin{abstract}

Spatial complementary code keying modulation (SCCKM) is proposed as a novel block coding modulation scheme. An input binary sequence is modulated based on the different lengths of complementary code keying (CCK) modulation and then spread across the transmit antennas (spatial domain) in a multiple input multiple output (MIMO) system exploiting orthogonal frequency division multiplexing (OFDM). At the receiver side, zero forcing equalization is applied to the OFDM modulated data to mitigate the effect of the multipath fast fading channel and then followed by maximum likelihood (ML) detection to retrieve the input sequence. The performance of SCCKM in different MIMO systems is compared to that of spatial modulation (SM) as a baseline scheme. Simulation results show that for the same spectral efficiency, SCCKM is able to substantially improve the bit error rate (BER).
\end{abstract}

\section{INTRODUCTION}

\IEEEPARstart{T}he increasing demand for a seamless wireless link with ubiquitous coverage and higher data rate is causing rapid growth of data traffic. Taking into account the limited RF spectrum, such data deluge motives to conduct further research to deliver more robust modulation, forward error coorection (FEC) and equalization techniques that can boost the performance over the physical layer. Significant progress on modulation and coding techniques have been achieved, many of which have been implemented in industry. However, despite of such progress, there is an inevitable need for new coding and modulation techniques that are more invulnerable to the channel impairments and interference.

Spatial modulation (SM) \cite{4382913}\cite{4176887}\cite{4149911} has been proposed as a modulation scheme which aggregates the spectral efficiency and mitigates the interference in a MIMO system. The scheme is based on simultaneous use of modulation symbol and transmit antenna number index as specific combinations to form sets of codes that will be distributed across the antennas \cite{4149911}. SM maps the data stream to unique combinations of a constellation symbol (according to modulation scheme) and an index number of one transmit antenna as two data units to carry the information, allowing an implicit transmission of $n$ extra bits per OFDM sub-channel where $n$ equals $\log_2 N_{t}$ and $N_{t}$ is the number of transmit antennas. Additional spatial multiplexing gain, no need for transmit antenna synchronization, and avoiding inter channel interference (ICI) are its key achievements.

\begin{figure}[t]
\centering
\includegraphics[scale=0.54]{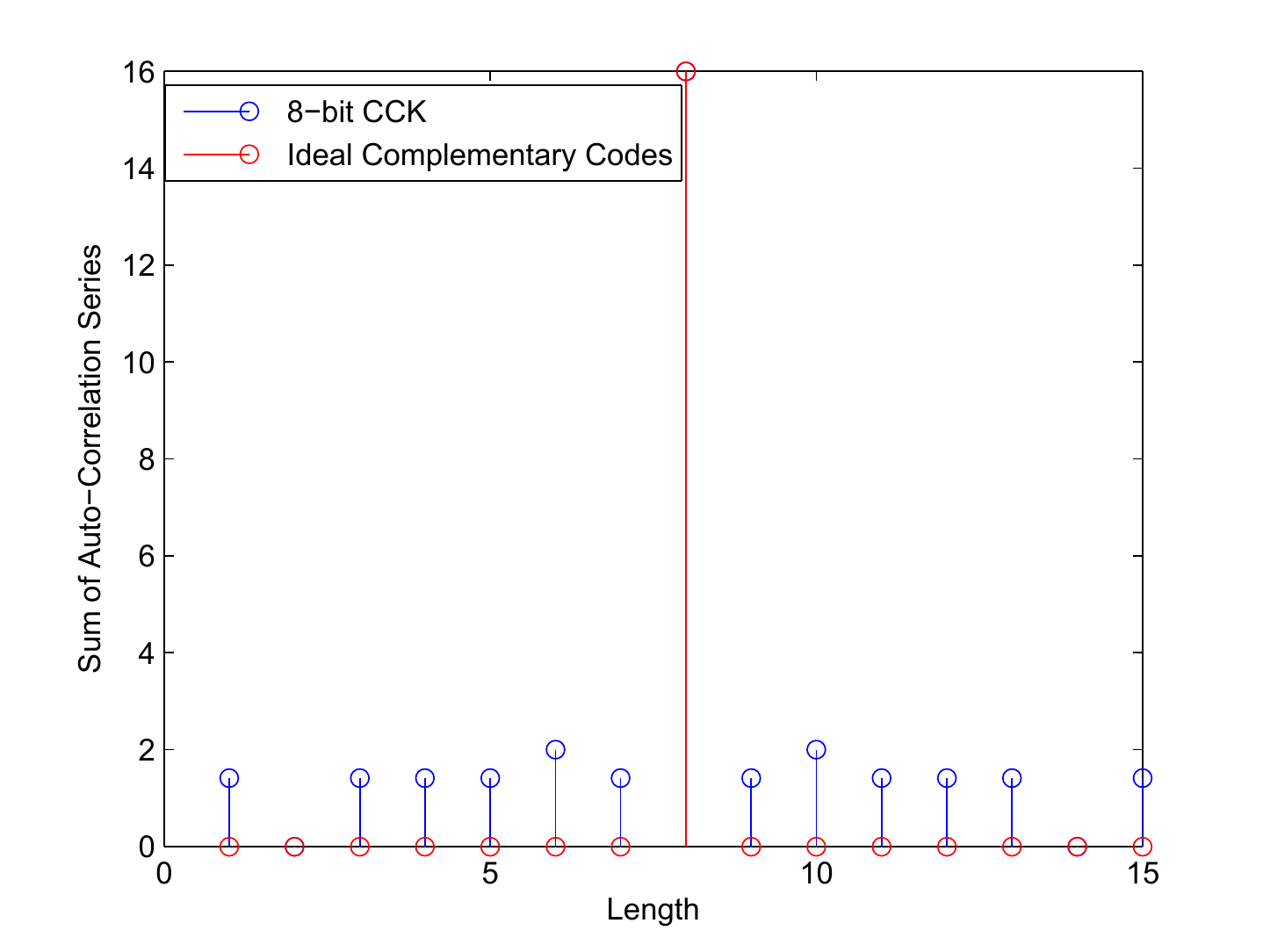}
\caption{Correlation properties of CCK codes.}
\label{fig:correlationprop}
\vspace{-4mm}
\end{figure}

Proposing more robust and spectrally efficient transmission techniques for MIMO systems, spatial complementary code keying modulation (SCCKM) is an appealing scheme by acquiring the characteristics of complementary sequences. Complementary sequences \cite{1057620} are promising set of codes that profit from impulse response shaped auto-correlation function. Indeed, the summation of the auto-correlation functions of any two complementary sequences for any non-zero shift is zero, enhancing the system's resistance against inter-symbol-interference (ISI). Complementary sequences establish the basis of complementary code keying (CCK) modulation which was first introduced as an 8-bit (chip) spreading sequence representing one symbol in IEEE802.11b \cite{6178212}\cite{4027571} offering a chip rate of 11 Mchps, and has been able to increase the data rate to 5.5Mbps and 11Mbps in an spread spectrum system \cite{4017717}. Fig. \ref{fig:correlationprop} shows the correlation properties of both complementary and 8-bit CCK codes. For further study on CCK modulation can refer to \cite{780033} \cite{1344577} \cite{4381105}. CCK codewords can be of various lengths, however, we only study length two, four and eight, leaving other lengths for future study. We further introduce a novel spatial modulation scheme by applying CCK modulation to the spatial domain known as SCCKM where CCK codewords are embedded across transmit antennas in a MIMO system. 

The rest of this paper is organized in the following manner. In Section \ref{sec:formulation}, we concisely discuss complementary and polyphase complementary sequences and then derive length two, four and eight CCK codes and discuss their properties. In section \ref{sec:spatial}, we present SCCKM and illustrate how CCK modulation is applied to a MIMO-OFDM system. In Section \ref{sec:simulation}, the performance of a SCCKM is studied and compared to that of SM. Conclusions are drawn in Section \ref{sec:conclusion}.

\vspace{-3.5mm}
\section{COMPLEMENTARY CODE KEYING MODULATION}
\label{sec:formulation}

\subsection{Complementary and Polyphase Complementary Sequences} 
\label{subsec:back}

Complementary sequences \cite{1057620} as core of CCK modulation, are set of codes for which the sidelobes of auto-correlation functions of any two complementary pair summate to zero. They consist of non-complex elements and are generated based on the following kernel. Considering the two primary sequences $M_{1}$ and $N_{1}$ representing ${1}$ and ${1}$, any longer sequence can be constructed using $M_{k}$ = $N_{k-1}$ $N_{k-1}$ and $N_{k}$ = $M_{k-1}$ $\hat N_{k-1}$ where $\hat N$ denotes the complement of the sequence $N$. Polyphase complementary sequences \cite{1056272} are similar with the difference that they are made up of complex elements \cite{594459}. Elements of a polyphase sequence have unit magnitudes with associated phase of $\theta$ and generated as presented by (\ref{eq:pol}). It is worth noting that phase recognition and accordingly (\ref{eq:pol}) play central role in generation of CCK codewords \cite{594459}. This will be realized in the sequel as generation of different lengths of CCK modulation are discussed.

\vspace{-3mm}

\begin{equation}
x(t) = \rm p(\textit {t}- \textit {k}T) e^{j\theta} \hspace{2mm} \rm and \hspace{2mm} p(\textit {t}) =\begin{cases}
     1 & \text{if $0<t<\rm T$} \\
     0 & \text{otherwise}
  \end{cases}
  \label{eq:pol}
\end{equation}

\vspace{-4mm}

\subsection{Two Bit CCK Modulation}

The 2-bit CCK modulation takes a sequence of $2$ bits as the input stream. Two bits refer to two phases as shown in matrix (\ref{eq:euler2}) and for sake of symmetry, they are determined as (${0,\pi}$). There are four phase combinations which are exploited by (\ref{eq:euler2}) to derive the four 2-bit CCK codewords, noting that each phase combination corresponds to one CCK codeword. Table.\ref{tab:cck2} shows the computed 2-bit CCK codwords.

\begin{equation}
\mathbf{M_{0}} = \left[ \begin{array}{cc} \phi_{1} & \phi_{1} \\ \phi_{2} & 0  \end{array}\right] \quad, \quad \mathbf{C_{0}} = \left[ {e^{j(\phi_{1}+\phi_{2})} , e^{j(\phi_{1})}} \right] 
  \label{eq:euler2}
\end{equation}

\begin{table}[b]
\renewcommand{\arraystretch}{2}
\centering
\caption{2-bit CCK Modulation}
\begin{tabular}{|c|c|}
\hline \textbf{Binary Sequence} & \textbf{CCK Codeword} \\ 
\hline 00 & 1,1 \\ 
\hline 01 & -1,1 \\ 
\hline 10 & -1,-1 \\ 
\hline 11 & 1,-1 \\ 
\hline 
\end{tabular}
\label{tab:cck2}
\end{table}

\vspace{-3mm}

\subsection{Four Bit CCK Modulation}
The 4-bit CCK modulation takes 4 bit sequences as the input data stream. The matrix shown in (\ref{eq:mat}) represents the phase arrangement associated with 4-bit CCK codewords which involves three phases. To maintain the symmetrical nature of the modulation, the three phases are evenly distributed in the constellation diagram and are specified as $(0,2\pi/3,4\pi/3)$. (\ref{eq:euler4}) is then used to obtain the 4-bit CCK codewords and demonstrates how the three phases presented in (\ref{eq:mat}), are fundamental to 4-bit CCK modulation.

\begin{equation}
\mathbf{M_{1}} = \left[ \begin{array}{cccc} \phi_{1} & \phi_{1} & \phi_{1} & \phi_{1} \\ \phi_{2} & 0 & \phi_{2} & 0 \\ \phi_{3} & \phi_{3} & 0 & 0 \end{array} \right]
\label{eq:mat}
\end{equation}

Considering the three phases, there are twenty seven phase combinations and accordingly twenty seven 4-bit CCK codewords computed by (\ref{eq:euler4}).

\vspace{-2mm}

\begin{equation}
	\mathbf{C_{1}} = \left[{e^{j(\phi_{1}+\phi_{2}+\phi_{3})} , e^{j(\phi_{1}+\phi_{3})} , e^{j(\phi_{1}+\phi_{2})} , -e^{j(\phi_{1})}}\right]
	\label{eq:euler4}
\end{equation}
 
To maintain the one-to-one correspondence between the possible sixteen $4$ bit binary input data streams and the generated 4-bit CCK codewords, the optimum subset containing only sixteen 4-bit CCK codewords is specified in three stages. In first stage, an statistically large number of random subsets where each contains sixteen 4-bit CCK codewords are selected. In second stage, the four dimensional Euclidean distance between all sixteen pairs of all subsets are computed. This stage narrows down the number of appropriate subsets by choosing the subsets that maximize the minimum Euclidean distance. In the last stage, the number of times that the specified minimum Euclidean distance exists within all remained subsets is counted and the subset that contains the least number of the specified minimum Euclidean distance between its pairs is selected as the sub-optimum 4-bit CCK codeword as represented in (\ref{eq:cckcode}). 

\begin{equation}
\left[ \begin{array}{cccc} 1 & 1 & 1 & -1 \\ 1 & -0.5+0.866i & -0.5-0.866i & -1 \\ -0.5+0.866i & -0.5+0.886i & 1 & -1 \\ -0.5-0.866i & -0.5-0.866i & 1 & -1 \\ -0.5-0.866i & 1 & -0.5-0.866i & -1 \\ -0.5-0.866i & -0.5+0.866i & -0.5+0.866i & -1 \\ -0.5+0.866i & -0.5-0.866i & -0.5-0.866i & -1 \\ 1 & -0.5-0.866i & -0.5-0.866i & 0.5-0.866i \\ 1 & 1 & -0.5+0.866i & 0.5-0.866i \\ -0.5+0.866i & 1 & -0.5-0.866i & 0.5-0.866i \\ -0.5-0.866i & -0.5+0.866i & -0.5-0.866i & 0.5-0.866i \\ -0.5+0.866i & -0.5+0.866i & -0.5+0.866i & 0.5-0.866i \\ 1 & -0.5-0.866i & 1 & 0.5+0.866i \\ -0.5+0.866i & -0.5-0.866i & -0.5+0.866i & 0.5+0.866i \\ -0.5-0.866i & 1 & -0.5+0.866i & 0.5+0.866i \\ -0.5+0.866i & 1 & 1 & 0.5+0.866i \end{array} \right]
\label{eq:cckcode}
\end{equation}

\vspace{2mm}

The input data stream will be divided into streams of 4 bits and then arbitrarily modulated to 4-bit CCK codewords presented in (\ref{eq:cckcode}) which will be spatially distributed across antennas.

\vspace{-3mm}

\subsection{Eight Bit CCK Modulation}
The 8-bit CCK modulation has been previously implemented in IEEE802.11b \cite{4027571}. The 8-bit CCK modulation splits the 8-bit stream ${a_{0}a_{1}a_{2}a_{3}a_{4}a_{5}a_{6}a_{7}}$ to four 2-bit sub-streams where each sub-stream is allocated a phase as presented in table \ref{tab:cck81}. Based on the binary 2-bit sub-stream, the corresponding phase of 2-bit sub-stream varies as shown in table \ref{tab:cck8}.

\begin{table}[t]
\renewcommand{\arraystretch}{2}
\centering
\caption{Sub-stream phase allocation in 8-bit CCK Modulation}
\begin{tabular}{|c|c|c|c|c|}
\hline \textbf{Bit Sequence} & $a_{1}a_{0}$ & $a_{3}a_{2}$ & $a_{5}a_{4}$ & $a_{7}a_{6}$ \\
\hline \textbf{Phase} & $\phi_{1}$ & $\phi_{2}$ & $\phi_{3}$ & $\phi_{4}$ \\
\hline 
\end{tabular}
\label{tab:cck81}
\vspace{-4mm}
\end{table} 

The four phases associated with 8-bit sequence are formulated in the matrix shown in (\ref{eq:eulermatrix}) which clearly demonstrates how the elements of (\ref{eq:euler}) are constructed \cite{594459}. Inspection of (\ref{eq:euler}) demonstrates that each of its elements is indeed an exponent that is powered to the sum of the phases forming columns of the matrix given in (\ref{eq:eulermatrix})  \cite{594459}. The minus sign of fourth and seventh elements follows the rule discussed in generation of complementary sequences. The 8-bit CCK modulated codewords are shown in table \ref{tab:cck83}.

\vspace{-2.5mm}

\begin{table}[b]
\renewcommand{\arraystretch}{2}
\centering
\caption{Numerical phase allocation in 8-bit CCK Modulation}
\begin{tabular}{|c|c|c|c|c|}
\hline $a_{k+1}a_{a}$ & 00 & 01 & 10 & 11 \\
\hline \textbf{Phase} & 0 & $\pi$ & $\pi/2$ & $-\pi/2$ \\
\hline 
\end{tabular}
\label{tab:cck8}
\end{table}

\begin{equation}
\mathbf{M_{2}} = \left[ \begin{array}{cccccccc} \phi_{1} & \phi_{1} & \phi_{1} & \phi_{1} & \phi_{1} & \phi_{1} & \phi_{1} & \phi_{1} \\ \phi_{2} & 0 & \phi_{2} & 0 & \phi_{2} & 0 & \phi_{2} & 0 \\ \phi_{3} & \phi_{3} & 0 & 0 & \phi_{3} & \phi_{3} & 0 & 0 \\ \phi_{4} & \phi_{4} & \phi_{4} & \phi_{4} & 0 & 0 & 0 & 0  \end{array}\right]
\label{eq:eulermatrix}
\vspace{-2mm}
\end{equation}

\begin{equation}
\begin{split}
\mathbf{C_{2}} =& \left[ e^{j(\phi_{1}+\phi_{2}+\phi_{3}+\phi_{4})}, e^{j(\phi_{1}+\phi_{3}+\phi_{4})}, e^{j(\phi_{1}+\phi_{2}+\phi_{4})}, \right. \\
  & \left. -e^{j(\phi_{1}+\phi_{4})}, e^{j(\phi_{1}+\phi_{2}+\phi_{3})}, e^{j(\phi_{1}+\phi_{3})}, \right. \\ & \left. -e^{j(\phi_{1}+\phi_{2})}, e^{j(\phi_{1})} \right]
\end{split}
\label{eq:euler}
\end{equation}
\vspace{0.03cm}

8-bit CCK modulation generates 256 different codewords where 64 codewords are distinctively orthogonal. This orthogonality property results in low cross-correlation and consequently efficient interference mitigation which allows the simultaneous transmission by multiple antennas in a MIMO system.

\vspace{-0.4cm}

\subsection{Minimum Distance Between Complementary Codes}

As discussed earlier, in order to minimize the bit error probability, it is required to maximize the Euclidean distance which highly impacts the performance of CCK modulation \cite{594459}. In following, the minimum Euclidean distance achieved by different CCK modulations considering their lengths and associated phases are derived.
\begin{table}[t]
\renewcommand{\arraystretch}{2.2}
\centering
\caption{8-bit CCK Modulation}
\begin{tabular}{|c|c|}
\hline 8-bit Binary Data Stream & {00111011} \\ 
\hline 2-bit Sub-Stream & {00}, {11}, {10}, {11} \\ 
\hline Phase Allocation & {0}, {-$\pi$/2}, {$\pi$/2}, {-$\pi$/2} \\ 
\hline 8-bit CCK Modulated Codeword & {-\textit{i},1,-1,\textit{i},1,\textit{i},\textit{i},1} \\ 
\hline 
\end{tabular}
\label{tab:cck83}
\vspace{-1.5mm}
\end{table}

For a complementary code of length $N$, there will be additional $\log_2 N$ orthogonal codes together forming a subset of length 1+$\log_2 N$. Considering there are $M$ phases associated with each code, the number of bits per codeword will be (1+ $\log_2 N$) $\times$ $\log_2 M$ and hence $1+ N/2$ symbols suffice to obtain the 1+$\log_2 N$ required number of phases through 1+$\log_2 N$ phase equations where each phase is used in $N/2$ of phase equations. The minimum Euclidean distance between any two codewords is given in (\ref{eq:dis}) considering that the least phasor rotation between $N/2$ symbols is $2\pi/M$. Having computated the Euclidean distance for all 2-bit ,4-bit and 8-bit CCK codwords, it is perceived that 8-bit CCK modulation profits from maximum Euclidean distance among its corresponding codewords in comparison to 2-bit and 4-bit CCK modulations. This considerably lowers the bit error probability which consequently improves the data rate.
\vspace{-1.4mm}
\begin{equation}
d_{min} = \sqrt{\frac{N}{2}[1-e^{j(\frac{2\pi}{M})}]}
\label{eq:dis}
\end{equation}

\section{Spatial Complementary Code Keying Modulation}
\label{sec:spatial}

\begin{figure*}[t]
\centering
\includegraphics[scale=1]{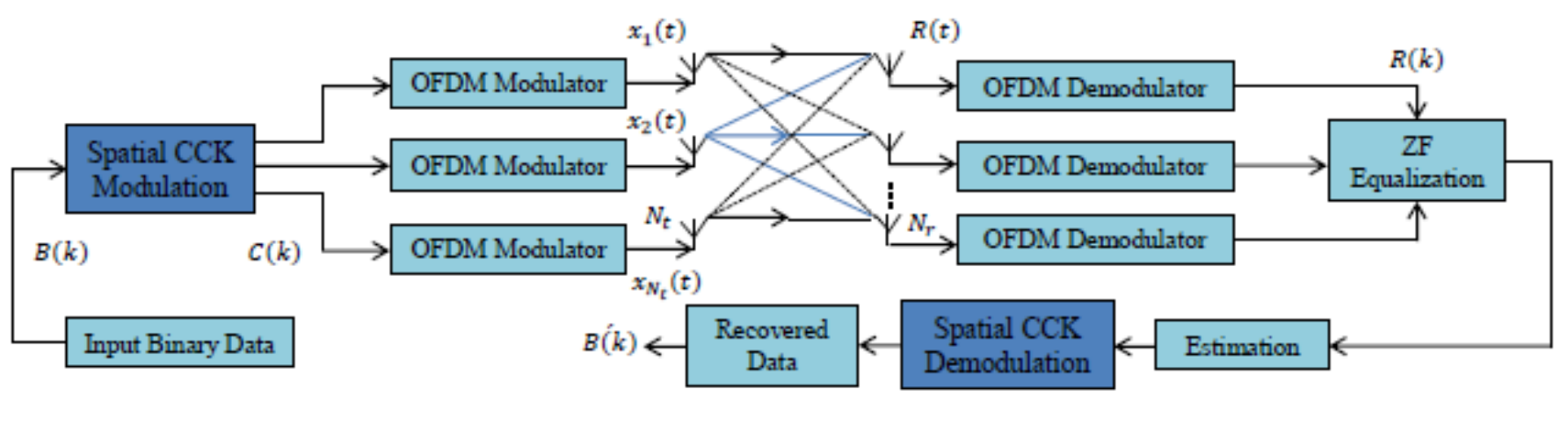}
\caption{Spatial complementary code keying modulation block diagram.}
\label{fig:cckblock}
\vspace{-0.4cm}
\end{figure*}

The matrix $B(k)$ of size $m \times N_{sub}$ denotes the input binary sequence intended for transmission through the channel where $m$ varies according to the order of CCK modulation and $N_{sub}$ refers to the number of OFDM sub-channels. CCK modulation is applied to the columns of $B(k)$ where each column will be block coded to generate CCK codewords of accordingly desired length (two, four and eight). The generated CCK codewords will be spatially spread across the transmit antennas proposing the SCCKM scheme. The matrix $C(k)$ is the output of SCCKM which is of dimension $N_{t}\times N_{sub}$ and $N_{t}$ is the number of transmit antennas. A bank of OFDM modulators are placed at the transmitter side where OFDM is applied to each row of $C(k)$. The OFDM modulated vectors will be simultaneously transmitted by $N_{t}$ transmit antennas across the time-variant wireless channel where additive white Gaussian noise and fading will degrade the signal.

The channel is represented by the block matrix $\textbf{H}(\tau,t)$ which is of dimension $N_{r}\times N_{t}$ and $N_{r}$ is the number of receive antennas. Each channel element is a vector of size $1\times p$ and $p$ represents the number of channel coefficients between each transmit-receive pair. 

\begin{equation}
	\textbf{H}(\tau,t)=\left(\begin{matrix}
								h_{1,1}(\tau,t) \quad \quad h_{1,2}(\tau,t) \quad \quad \cdots \quad \quad h_{1,N_{t}}(\tau,t)\\
								h_{2,1}(\tau,t) \quad \quad h_{2,2}(\tau,t) \quad \quad \cdots \quad \quad h_{2,N_{t}}(\tau,t)\\
								\vdots \quad \quad \quad \quad \vdots \quad \quad \quad \ddots \quad \quad \quad \vdots \\
								h_{N_{r},1}(\tau,t) \quad \quad h_{N_{r},2}(\tau,t) \quad \quad \cdots \quad h_{N_{r},N_{t}}(\tau,t)\\
							\end{matrix}
				\right)
	\label{eq:partition_Hti}
\end{equation}

\vspace{-0.2cm}

\begin{equation}
h_{m,n}(\tau,t)=\left[h_{m,n}(\tau,t)^1 h_{m,n}(\tau,t)^2 \quad \cdots\quad h_{m,n}(\tau,t)^p \right]
\label{eq:chpath}
\end{equation}

\vspace{0.4cm}

OFDM demodulation will be applied to rows of the received signal matrix $\textbf{R}(t)$ using the $N_{r}$ demodulators at the receiver. The output from OFDM demodulation $R(k)$ is a matrix of size $N_{r}\times m$. To lessen the effect of the channel impairments, the output from OFDM demodulation is equalized in the frequency domain by applying the zero forcing (ZF) equalization to each of its column vectors (More effective equalization techniques like minimum mean square (MMS) are left for future study). Note that $k$ represents the discrete time index. The equalization process is performed on a sub-channel basis which requires the channel corresponding to each sub-channel. It should be mentioned that for non-symmetrical MIMO systems, the pseudo-inverse is implemented for ZF equalization. The equalized data stream will be divided into sub-streams received on each sub-channel, and the Euclidean distance between the sub-stream received on each sub-channel and contents of the CCK codebook (which consists of all the generated CCK codewords) will be computed. In this scenario, the CCK codebook consists of $L=m^2$ CCK codewords. The codeword $E(k)$ resulting in minimum Euclidean distance will be opted as desired one. The opted codeword will then be de-mapped to the corresponding data sequence. The cross-correlation properties of CCK codes mitigate the impact of inter-antenna interference and dismiss the need for inter-antenna synchronization. This allows simultaneous transmission by all antennas on each subcarrier and hence achieving spatial multiplexing gain.

\vspace{-3mm}
\begin{equation}
	\textbf{R}(t)=\textbf{H}(\tau,t)\otimes\textbf{X}(t)+ \textbf{N}(t)
	\label{eq:noise}
\end{equation}

	\vspace{-0.3cm}

\begin{equation}
	z(k)=\textbf{H}(k)^{-1} r(k)
\end{equation}

$\textbf{X}(t)$ and $\textbf{N}(t)$ present the matrix of OFDM symbols and the matrix of the additive white Gaussian noise (AWGN), respectively and $\otimes$ denotes convolution in time domain.

\vspace{-3mm}
\begin{eqnarray}
{d_i}^2(\textbf{c}_j,\textbf{s}_i)=(\textbf{c}_j-\textbf{s}_i)(\textbf{c}_j^*-\textbf{s}_i^*) \nonumber \\ \centering i=1,2,...,N_{sub} \hspace{0.6cm} j=1,2,...,L
\end{eqnarray}

\vspace{-0.55cm}

\begin{equation}
E_i(k)=min({d_i^2})
\vspace{-1mm}
\end{equation}

${d}^2(\textbf{c}_j,\textbf{s}_i)$ refers to the squared Euclidean distance between the received signal $\textbf{s}_i$ on each sub-channel and each content of the CCK codebook $\textbf{c}_j$. $E_i$ presents the CCK codeword with minimum Euclidean distance associated with $i^{th}$ sub-channel.

\begin{table}[b]
\vspace{-2mm}
\centering
\renewcommand{\arraystretch}{1.4}
\caption{System Simulation Parameters}
\label{tab:par}
\scalebox{0.85}{
\centering
\begin{tabular}{|>{\centering\arraybackslash}p{4.1cm}|>{\centering\arraybackslash}p{1.8cm}|}
\hline \textbf{Parameter} &  \textbf{Setting} \\ \hline
Carrier Frequency (GHz)             		& 2      \\ \hline
Transmission Bandwidth (MHz)             	& 20     \\ \hline
Number of sub-channels       				& 256      \\ \hline
Number of Frames                 		& 1000     \\ \hline
Numer of OFDM symbols     		& 20      \\ \hline
Symbol Period (ns)              		& 50     \\ \hline
Cyclic Prefix Length           	& 16     \\ \hline
Channel Delay Spread (ns)            		& 50     \\ \hline
Transmit Power (W)                			& 1     \\ \hline
\end{tabular}}
\vspace{-5mm}
\end{table}

\vspace{-1mm}
\section{SIMULATION}\label{sec:simulation}

We consider different MIMO systems and transmit $1000$ frames where each contains $20$ OFDM symbols. The channel is assumed to be a time-variant frequency-selective multipath channel with a maximum delay spread of $50$ \textit{ns}. The multipath channels are assumed to be statistically independent. The noise is a temporally and spatially additive white Gaussian one. It is assumed that full knowledge of the channel is available at the receiver with perfect time and frequency synchronization. The normalized signal-to-interference-plus-noise ratio (EbN0) is used. Table. \ref{tab:par} summarizes the simulation parameters.

\begin{figure}[t]
\centering
\includegraphics[scale=0.53]{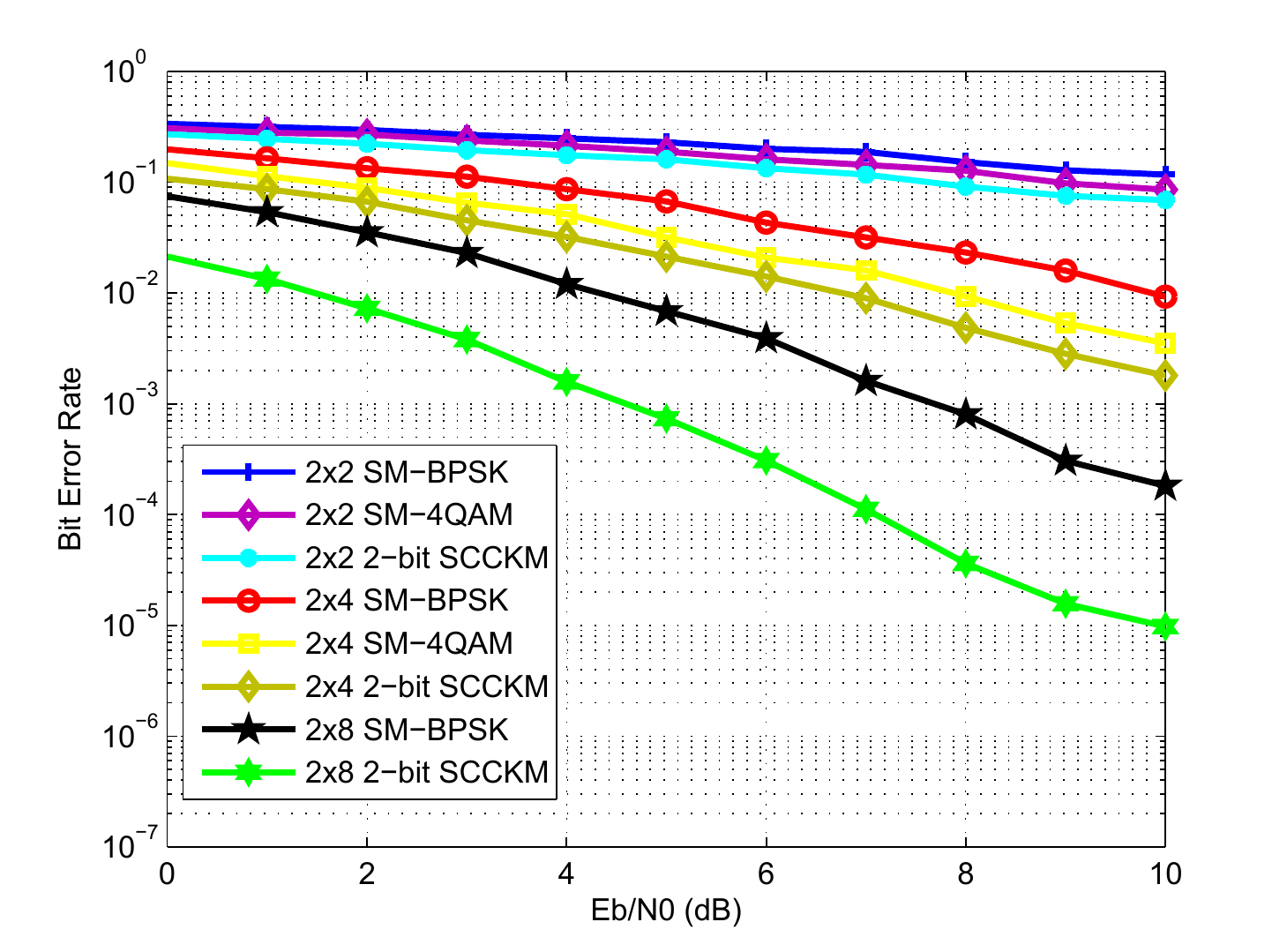}
\caption{BER comparison of 2-bit SCCKM and SM.}
\label{fig:cck2}
\vspace{-6mm}
\end{figure}

Fig.\ref{fig:cck2} illustrates the performances of 2-bit SCCKM and SM. The number of transmit antennas is kept at $2$ while using $2,4$ and $8$ antennas at the receiver. It is realized that spatial diversity at the receiver side has a substantial impact on the performance of both schemes. However, this impact is further boosted in SCCKM. At EbN0 of 10 dB, increasing the number of receive antennas from 2 to 4 and 4 to 8 lowers the BER by $\sim$ 38.46x and 183.59x in SCCKM and $\sim$ 12.65x and 51.02x in SM-BPSK. In a 2x8 MIMO, the starting BER of SCCKM is also $\sim$ 3.52x lower than SM-BPSK.

\begin{figure}
\centering
\includegraphics[scale=0.53]{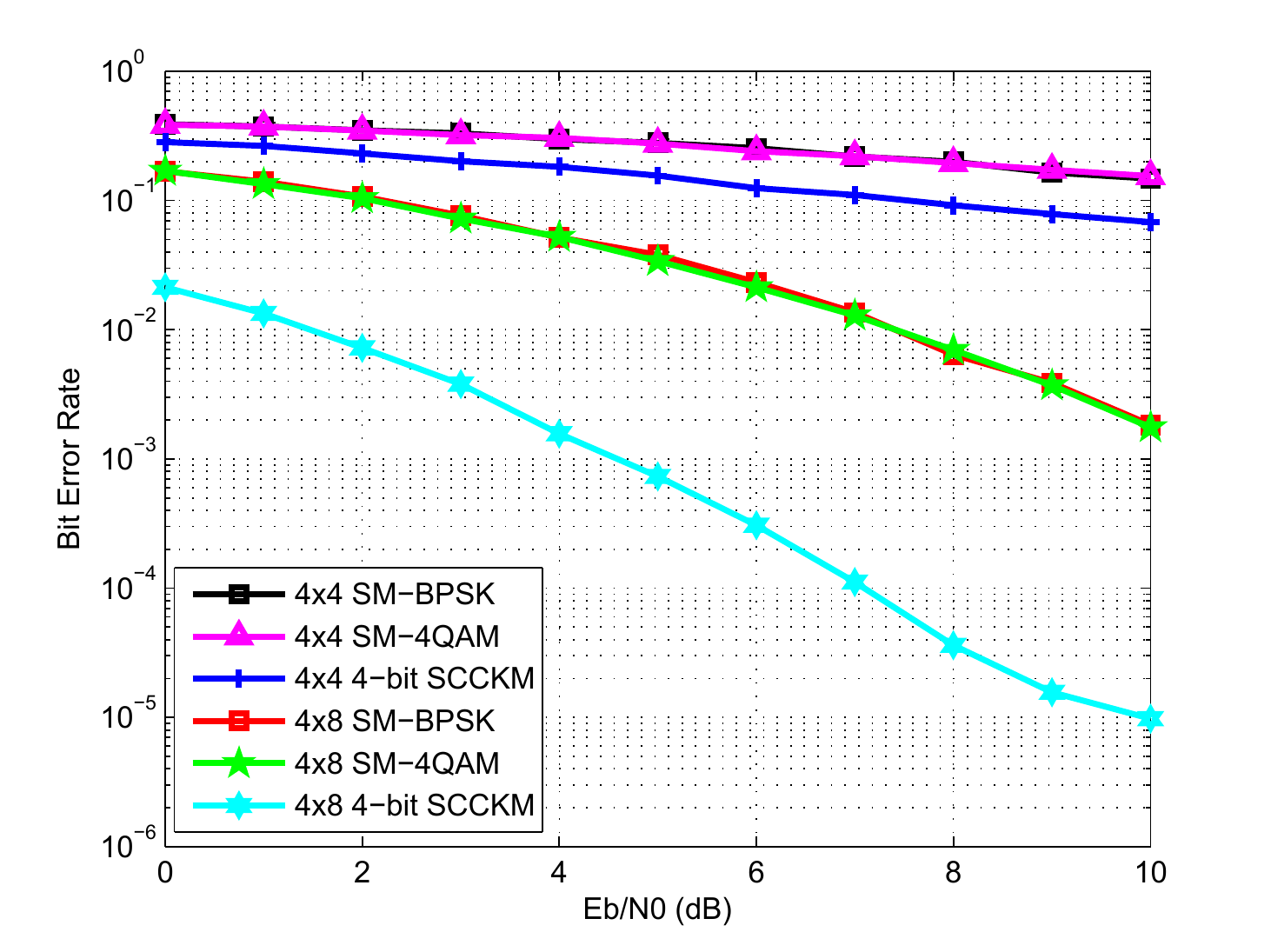}
\caption{BER comparison of 4-bit SCCKM and SM.}
\label{fig:cck4}
\vspace{-6mm}
\end{figure}

Fig.\ref{fig:cck4} shows that in a 4x4 MIMO (employing the same number of antennas at both transmitter and receiver sides), 4-bit SCCKM outperforms the SM. The SCCKM has a higher spectral efficiency as it transmits $4$ bits per sub-channel comparing to SM-BPSK which sends $3$ bits. SCCKM and SM-4QAM offer the same spectral efficiency. However, at EbN0 of 10 dB, the SCCKM is capable to achieve $\sim$ 1.6x lower BER than SM-4QAM. As pointed earlier, increasing the number of receive antennas leverages spatial diversity which would improve the performance of both SM and SCCKM schemes. Fig.\ref{fig:cck4} also illustrates the performances of SCCKM and SM in a 4x8 MIMO system. It is clearly seen that at EbN0 of $9$ dB, 4-bit SCCKM reaches the BER of $0.06 \times 10^{-4}$ in comparison to BER of $0.07 \times 10^{-2}$ achieved by SM-BPSK and SM-4QAM. This suggests that while SM offers a lower spectral efficieny (SM-BPSK transmits $3$ bits per sub-channel), it yet requires further increase in signal power to reach the same BER.
  
\begin{figure}[t]
\centering
\includegraphics[scale=0.53]{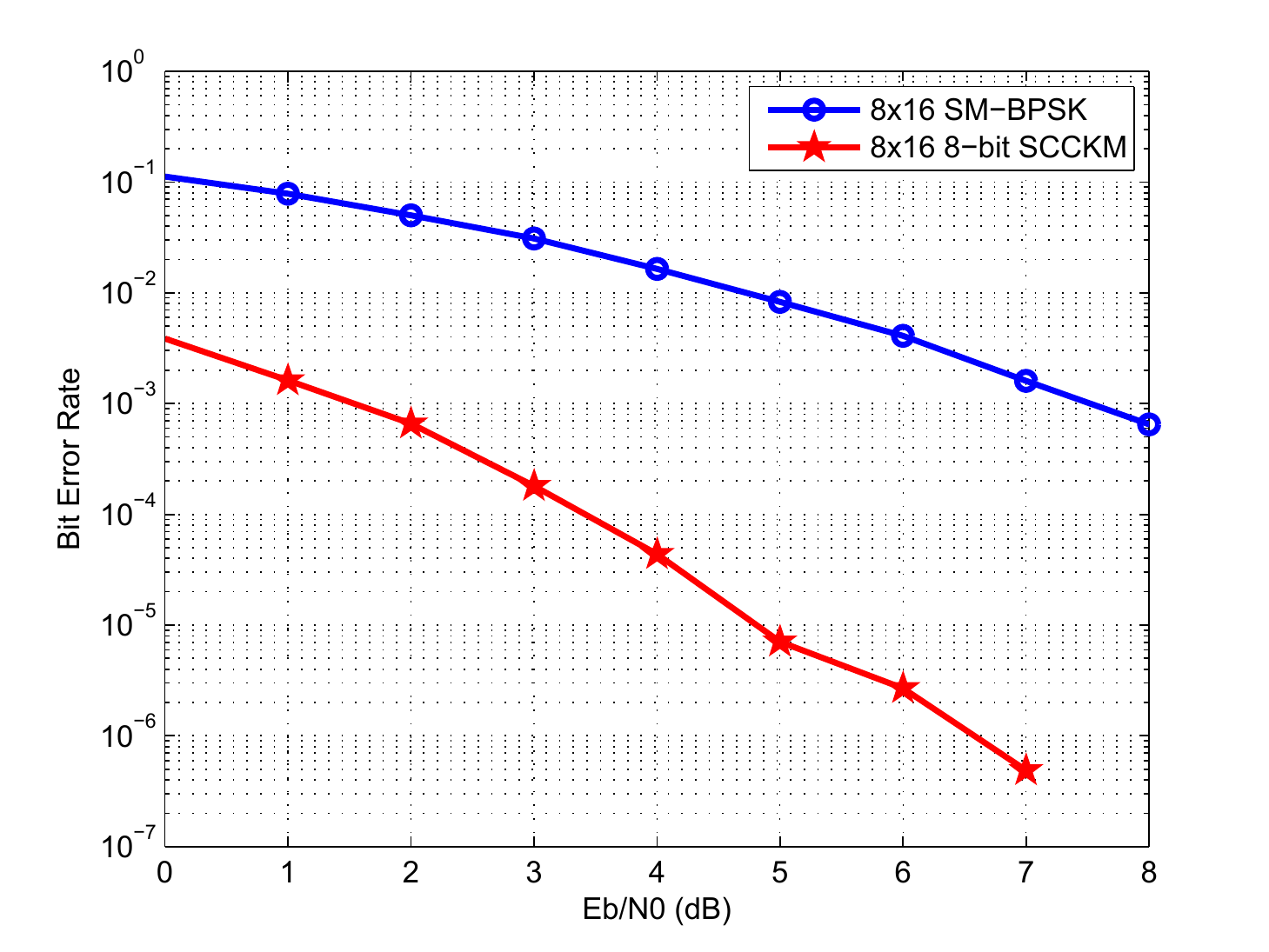}
\caption{BER comparison of 8-bit SCCKM and SM.}
\label{fig:cck8}
\vspace{-6mm}
\end{figure}

Fig.\ref{fig:cck8} shows the performance of 8-bit SCCKM in a 8x16 MIMO system. Achieving the BER of $10^{-7}$ at BER of less than 8 dB clearly demonstrates the unique performance of 8-bit SCCKM. SM-BPSK only achieves the BER of $10^{-3}$ at EbN0 of 8 dB. This performance is partly due to spatial diversity at the receiver using 16 antennas, and partly due to the point that 8-bit SCCKM has got the best correlation properties in terms of least cross-correlation comparing to 2-bit and 4-bit SCCKM. The low cross-correlation of CCK codes (and in particular 8-bit CCK), considerably boosts their resistivity to both channel impairments and inter-antenna interference.

\vspace{-2.5mm}

\section{CONCLUSION}\label{sec:conclusion}

SCCKM as a novel spatial modulation scheme based on different lengths of CCK modulation was proposed. It was expressed that CCK modulation profits from low cross-correlation property which enhances its resistance to interference. This particularly mitigates the inter antenna interference and eases antenna synchronization in MIMO systems where all antennas can actively transmit at the same time instant. SCCKM was applied to a MIMO-OFDM system and its performance was compared to that of SM. SCCKM outperforms the SM by considerably lowering the BER at the same EbN0. In a MIMO system with more antennas at the receiver side, SCCKM results in notable BER reduction with starting BER being less than $10^{-1}$. The 8-bit SCCKM was also realized as the most promising SSCKM due to its cross-correlation properties. 

\vspace{-3mm}

\bibliographystyle{ieeetr}
\bibliography{references}

\end{document}